\documentclass[aps,twocolumn,showpacs,citeautoscript,reprint,prb,floatfix]{revtex4-1}

\usepackage{bm}
\usepackage{graphicx}
\usepackage{amsmath}
\usepackage{amsfonts}
\usepackage{bbold}
\usepackage{amssymb}
\usepackage{siunitx}
\usepackage{color}
\newcommand{\ii}{\mathrm{i}}

\newcommand{\sgn}{\mathrm{sgn\,}}

\begin{document}

\title[Inducing anisotropies in Dirac fermions by periodic driving]{Inducing anisotropies in Dirac fermions by periodic driving}

\author{A. D\'{i}az-Fern\'{a}ndez}

\affiliation{GISC, Escuela T\'{e}cnica Superior de Ingeniería (ICAI), Universidad Pontificia Comillas, E-28015 Madrid, Spain}

\begin{abstract}

We consider the three-dimensional Hamiltonian for Bi$_2$Se$_3$, a second-generation topological insulator, under the effect of a periodic drive for both in-plane and out-of-plane fields. As it will be shown by means of high-frequency expansions up to second order in the Floquet Hamiltonian, the driving induces anisotropies in the Dirac cone and opens up a quasienergy gap for in-plane elliptically polarized fields. Analytic expressions are obtained for the renormalized velocities and the quasienergy gap. These expressions are then compared to numerical calculations performed by discretizing the Hamiltonian in a one-dimensional lattice and following a staggered fermion approach, achieving a remarkable agreement. We believe our work may have an impact on the transport properties of topological insulators.


\end{abstract}


\maketitle

\section{Introduction}

Three-dimensional topological insulators host an odd number of massless helical Dirac fermions at the surface which owe their existence to the non-trivial topology of the bulk Hilbert space~\cite{HasanKane10,HasanMoore11,QiZhang11}. A plethora of proposals exist towards exploiting these Dirac cones in quantum transport devices due to their unprecedented characteristics (see Ref.~\cite{Culcer12} for a review). As such, reshaping the linear spectrum becomes particularly interesting as it corresponds to modifying the Fermi velocity~\cite{Li09,Trambly10,Hwang12,Miao13,Lima16,Diaz-Fernandez17a}. In particular, anisotropies in the dispersion are becoming particularly relevant~\cite{Hirata16,Moon11,Rusponi10,Wang14,Park11} and it has been shown to have a direct impact on the conductance~\cite{Trescher15}. Another key feature of the Dirac cones is their protection by time-reversal symmetry. By breaking such a symmetry, the Dirac cones at the surface are gapped out and the three-dimensional topological insulator hosts chiral hinge modes, thereby becoming a higher-order topological insulator~\cite{Schindler18}. In other words, the surface of the three-dimensional topological insulator becomes a Chern insulator~\cite{Haldane88,QiWuZhang06}.

On a different front, Floquet engineering of topological phases leads to novel states which can be manipulated by external periodic drivings~\cite{Lindner11,Rechtsman13,Fleury16,Grifoni98,Platero04,GomezLeon13a, Cayssol13,Kitagawa11,Oka19,Rudner20}. Examples include quasienergy gaps and Dirac cone widening in graphene~\cite{Kitagawa11,Syzranov13,Usaj14,Agarwala16} and topological insulators~\cite{Kitagawa11,Yudin16}. This dynamic tuning of the spectrum proves to be necessary in view of the aforementioned effect on transport properties. Previous works on topological insulators consider surface effective Hamiltonians~\cite{Kitagawa11} and are therefore limited to in-plane configurations. As it has been recently shown in Ref.~\onlinecite{Diaz-Fernandez19}, out-of-plane fields also reshape the Dirac cones. However, in contrast to in-plane fields, hybridization with the bulk states becomes unavoidable, a fact that is not accounted for by surface effective Hamiltonians. In this paper, we propose to consider the full three-dimensional Hamiltonian to obtain a Floquet Hamiltonian up to second order in a high-frequency expansion. As it will be shown, the reshaping of the cones becomes apparent and analytic expressions for the velocities and the quasienergy gaps are obtained. These analytic expressions are compared to numerical calculations obtained by discretizing the Hamiltonian along the direction perpendicular to the interface between the topological insulator and a trivial insulator and using a staggered fermion approach~\cite{Montvay94}. The details of the numerical calculations will not be presented herein and can be found in Ref.~\onlinecite{Diaz-Fernandez19}. As it will be shown, the agreement between both approaches is noteworthy. We firmly believe that the results presented in this paper will pave the way towards new devices which will profit from the dynamic modulation of the Dirac cones. 

\section{Theoretical Model}

The model we shall use is a continuum low-energy Hamiltonian around the $\Gamma-$point for Bi$_2$Se$_3$. Such a model can be obtained by means of the theory of invariants and to lowest order it is corresponds to a (3+1)-Dirac equation for the envelope function~\cite{Zhang09}. This Hamiltonian possesses two topologically distinct insulating ground states which can be labeled by means of a $\mathbb{Z}_2$ topological index~\cite{Ryu10,Schnyder08,Chiu16}. Such index corresponds to the sign of the mass term in the Dirac Hamiltonian~\cite{Zhang12}. The bulk-boundary correspondence predicts the existence of surface states upon placing together two systems of different index, thereby creating a topological boundary~\cite{Zhang12}. 

In the orbital-spin basis, the low-energy description of Bi$_2$Se$_3$ is therefore given by
\begin{equation}
    \mathcal{H}=\bm{\alpha}\cdot\bm{p}+m\beta \ ,
    \label{eq:01}
\end{equation}
where $\alpha_i=\tau_xs_i$ and $\beta=\tau_z$ are the Dirac matrices, with $\tau_i$ and $s_i$ the Pauli matrices in the orbital and spin subspaces, respectively, and $\bm{p}$ is the momentum operator. We are working in dimensionless units where $m=\pm 1$ is half the bulk energy gap. In the bulk where there is continuous translation symmetry, momentum is a good quantum number and the quasiparticles are massive Dirac fermions
\begin{equation}
 E = \pm\sqrt{p^2+m^2} \ ,
    \label{eq:02}
\end{equation}
where $p=|\bm{p}|$. Therefore, the spectrum for $m$ and $-m$ is exactly the same. However, the insulating ground states corresponding to these two opposite energy gaps belong to different topological sectors, characterized by a $\mathbb{Z}_2$ topological index, $\nu=\sgn(m)$. In fact, this model belongs to the AII class as it possesses time-reversal symmetry squaring to minus one, which indeed predicts a $\mathbb{Z}_2$ index characterizing two topologically distinct phases~\cite{Ryu10}. The bulk-boundary correspondence states that if a topological boundary is considered, meaning a system comprising two materials of opposite index, there will be massless excitations at the boundary~\cite{Zhang12}. The simplest model to show that this is indeed the case is to consider a sharp boundary where $m$ only changes sign upon crossing the boundary, that is, $m=\sgn(z)$. In that case, the in-plane momentum $\bm{p}_{\bot}=(p_x,p_y,0)$ is still a good quantum number since there is translation symmetry along the $XY$ plane. There is also continuous rotation symmetry about the $Z$-axis, which implies that the spectrum must be isotropic and can only depend on $p_{\bot}=|\bm{p}_{\bot}|$. Thus, the Hamiltonian for a topological boundary is
\begin{equation}
 \mathcal{H}=\bm{\alpha}\cdot\bm{p}+\beta\sgn(z) \ ,
    \label{eq:03}
\end{equation}
so that $\mathcal{H}\Psi=E\Psi$. We want to find surface states localized at the boundary. Thus, we use as an \emph{ansatz} $\Psi=\exp(-\lambda|z|)|\Phi\rangle$ where $\lambda>0$ and $|\Phi\rangle$ is position-independent. If $\bm{p}_{\bot}=\bm{0}$, we expect from symmetry that the surface state will be at zero energy and $|\Phi\rangle$ satisfies
\begin{equation}
  \ii\alpha_z\beta |\Phi\rangle= \lambda |\Phi\rangle \ ,
    \label{eq:04}
\end{equation}
which implies that $\lambda^2=1$. Since $\lambda>0$, we must have $\lambda=1$, so $\Psi=\exp(-|z|)\Phi$. Thus, $\Phi$ is an eigenvector of $\ii\alpha_z\beta=\tau_ys_z$ with eigenvalue $+1$. There are two such doubly-degenerate eigenvectors
\begin{equation}
 |\Phi\rangle_{\pm}=\frac{1}{\sqrt{2}}|\pm\rangle_{y}|\pm\rangle_{z} \ ,
    \label{eq:05}
\end{equation}
where $\sigma_i|\pm\rangle_{i}=\pm|\pm\rangle_{i}$. These are related by the time-reversal symmetry operator $\mathcal{T}=\ii s_y\mathcal{K}$ where $\mathcal{K}$ denotes complex conjugation. Indeed, $ |\Phi\rangle_{-}=\mathcal{T} |\Phi\rangle_{+}$. In other words, $ |\Phi\rangle_{+}$ and $|\Phi\rangle_{-}$ form a Kramers' pair. 

The term $\bm{\alpha}_{\bot}\cdot\bm{p}_{\bot}$ breaks the degeneracy for $\bm{p}_{\bot}\neq\bm{0}$ by coupling $ |\Phi\rangle_{+}$ and $ |\Phi\rangle_{-}$. In fact, we can see that $\alpha_x|\Phi\rangle_{\pm}=\mp\ii|\Phi\rangle_{\mp}$ and $\alpha_y|\Phi\rangle_{\pm}=|\Phi\rangle_{\mp}$, which implies that the diagonal elements are zero, whereas the off-diagonal elements are simply $_{\mp}\langle\Phi|\alpha_x|\Phi\rangle_{\pm}=\mp \ii$ and $_{\mp}\langle\Phi|\alpha_y|\Phi\rangle_{\pm}=1$. Hence, the Hamiltonian in the subspace of the surface states will be
\begin{equation}
 \mathcal{H}_{\mathrm{S}} = \left(\bm{\sigma}\times\bm{p}_{\bot}\right)\cdot\hat{\bm{z}} \ ,
  \label{eq:06}
\end{equation}
which corresponds to a Rashba Hamiltonian with $\sigma_i$ being the Pauli matrices in the surface states' subspace. Because of the absence of a $\bm{p}_{\bot}^2$ term in the Hamiltonian, the spectrum is not that of Rashba but an isotropic Dirac cone instead, $E=\pm p_{\bot}$. The surface states show well-defined helicities or spin-momentum locking where $\langle\bm{\sigma}\rangle$ and $\bm{p}_{\bot}$ are orthogonal to each other. 

\section{Periodically driven topological boundary}

Consider a topological boundary described by the Dirac Hamiltonian of equation~(\ref{eq:03}) and irradiate it with a periodically driven electromagnetic wave. We work in a gauge where the electrostatic potential vanishes and consider small enough samples so that the vector potential is only a function of time and we consider it to be periodic with period $T$, $\bm{A}(t)=\bm{A}(t+T)$. Minimally coupling to this gauge field we have
\begin{equation}
 \mathcal{H} = \bm{\alpha}\cdot(\bm{p}+\bm{A})+\beta\sgn(z) \ .
    \label{eq:13}
\end{equation}
We shall consider for $\bm{A}(t)$ 
\begin{equation}
 \bm{A} = \bm{a}e^{\ii\omega t}+\bm{a}^{*}e^{-\ii\omega t} \ ,
    \label{eq:14}
\end{equation}
where $\bm{a}$ is a vector whose components are $a_j=(f_j/2\omega)\exp(\ii\theta_j)$ with $j=1,2,3$. Here $f_j$ is the amplitude of the $j$-th component of the electric field, the phases $\theta_j$ are included to allow for the study of different polarizations, and $\omega=2\pi/T$ is the driving frequency. Since the Hamiltonian is time-periodic we shall use Floquet theory to find the quasienergies of the system. As discussed in the appendix, a Floquet Hamiltonian can be introduced as a static version of the actual Hamiltonian under stroboscopic evolution of the system~\cite{Rudner20}. It must be carefully remembered that such a Floquet Hamiltonian is not the true Hamiltonian of the system, although for high frequencies it captures essentially the same physics as the Hamiltonian of the system~\cite{Rudner20}. Analytic expressions can be obtained for the Floquet Hamiltonian by performing high-frequency expansions~\cite{Bukov15,Eckardt15,Goldman14}. The relevant expressions are given in the appendix. Additionally, we will solve equation~(\ref{eq:07}) numerically by discretizing the $Z$-direction in a lattice and using staggered fermions to avoid fermion doubling~\cite{Montvay94}. We shall not discuss the details of the numerical calculations, for more information the reader is referenced to Ref.~\onlinecite{Diaz-Fernandez19}. The numerical results can be summarized as follows: (1) in-plane (i.e. parallel to the topological interface) linearly polarized fields lead to anisotropic Dirac cones, where the widening occurs in the direction perpendicular to the field amplitude; (2) in-plane circularly polarized fields gap out the Dirac cone and widen the cone isotropically; (3) out-of-plane (i.e. perpendicular to the topological interface) linearly polarized fields lead to an isotropic widening of the cone, similar to the static case~\cite{Diaz-Fernandez17a}; (4) out-of-plane circularly polarized fields lead to a combination of (1) and (3). Out-of-plane fields also lead to hybridization with the bulk states at quasienergies closer to the Floquet-Brillouin zone edges~\cite{Diaz-Fernandez19}. As it will be shown, the high-frequency expansions up to second order correctly captures the features observed in the numerical calculations, with only slight deviations in the case of out-of-plane circularly polarized fields. This is possibly due to the aforementioned hybridizations, which become more prominent as the field is increased.

\begin{figure}[htb]
\centerline{\includegraphics[width=0.95\columnwidth]{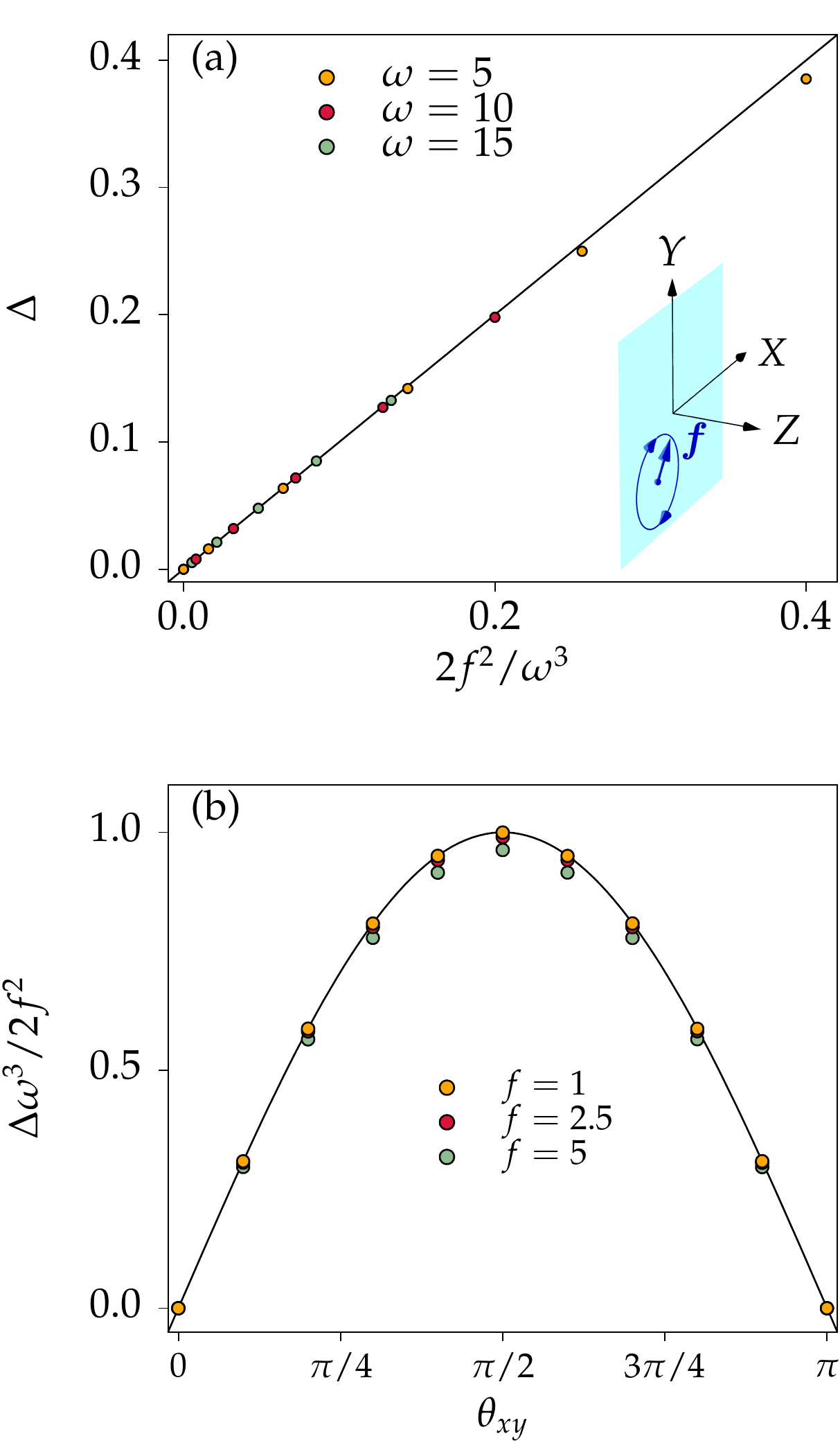}}
\caption{Quasienergy gap openings for in-plane fields with $f_x=f_y\equiv f$. The dots represent the numerical solution to the full problem and the black lines correspond to equation~(\ref{eq:tbreak_04}). (a) Circularly polarized field for different frequencies and (b) changing polarization for $\omega=5$ and different values of $f$.}
\label{fig:gap}
\end{figure}

In order to perform high-frequency expansions of the Floquet Hamiltonian we need the Fourier components of the true Hamiltonian, equation~(\ref{eq:07}). The only non-zero Fourier components are
\begin{equation}
 \mathcal{H}_0 = \bm{\alpha}\cdot\bm{p} + \beta\sgn(z) \ , \quad  \mathcal{H}_{-1} = \mathcal{H}^{\dagger}_{1}=\bm{\alpha}\cdot\bm{a} \ .
\label{eq:15}
\end{equation}
Since the only non-zero components are $m=0,\pm 1$, the high-frequency expansion [see equation~(\ref{eq:12})] simply reads
\begin{equation}
  \mathcal{H}_{\mathrm{F}}\simeq \mathcal{H}_0+\frac{1}{\omega}\left[\mathcal{H}_1,\mathcal{H}_{-1}\right]+\frac{1}{2\omega^2}\left(\left[\mathcal{H}_{-1},[\mathcal{H}_0,\mathcal{H}_1]\right]+\mathrm{h.c.}\right)
  \label{eq:16}
\end{equation}
Evidently, the first order approximation cannot lead to a widening of the Dirac cone since it only couples $\bm{p}$-independent terms. Therefore, it can only lead to gap openings. This is why, in order to explore the widening, it becomes necessary to push the expansion up to second order. After some tedious algebra, the first order term reads
\begin{equation}
    \delta\mathcal{H}^{1}= \frac{1}{2\omega^3}\sum_{ijk}\epsilon_{ijk}s_if_jf_k\sin\theta_{jk} \ ,
    \label{eq:19}
\end{equation}
where $\epsilon_{ijk}$ is the Levi-Civita symbol and we have defined $\theta_{ij}=\theta_i-\theta_j$ and $\bm{f}=(f_x,f_y,f_z)$. Before diving into the second order term, let us explore the possible gap openings that can occur due to $\delta\mathcal{H}^1$. As we can immediately see, such a term breaks time-reversal symmetry whenever the polarization is not linear, meaning that $\theta_{ij}\neq 0$ for some pair $i\neq j$. Indeed, if the polarization is linear, we immediately have $\delta\mathcal{H}
^1=0$. In all other cases, we can see the breaking of time-reversal symmetry
\begin{equation}
 \mathcal{T}\delta\mathcal{H}^1\mathcal{T}^{-1} = -\delta\mathcal{H}^1 \ .
 \label{eq:tbreak_01}
\end{equation}
However, it turns out that such a term only opens a quasienergy gap for the surface states if the field has two non-zero in-plane components with a non-zero phase difference $\theta_{ij}$ between two such components. Indeed, up to first order, we have that $\mathcal{H}_{\mathrm{F}}\simeq\mathcal{H}_0+\delta\mathcal{H}^1$, so we may project $\delta\mathcal{H}^{1}$ on to the subspace of the surface states given in equation~(\ref{eq:05}). Notice that we need not worry about the actual $z$- dependence of the bispinor through $\exp(-|z|)$ since $\delta\mathcal{H}^1$ is position independent. Moreover, the off-diagonal terms are zero since $\delta\mathcal{H}^{1}$ is $\tau$-independent. Thus, $_{\mp}\langle\Phi|\delta\mathcal{H}^{1}|\Phi\rangle_{\pm}=0$. As a result, $\delta\mathcal{H}^1$ does not couple $|\Phi\rangle_{\pm}$ and $|\Phi\rangle_{\mp}$, which implies that it can only affect at $\bm{p}_{\bot}=\bm{0}$, thereby possibly opening quasienergy gaps. Since $_{z}\langle\pm|\sigma_i|\pm\rangle_{z}=\pm\delta_{iz}$ and $_{y}\langle\pm|\tau_0|\pm\rangle_{y}=2$, then
\begin{equation}
    _{\pm}\langle\Phi|\delta\mathcal{H}^1|\Phi\rangle_{\pm}=\pm\frac{f_xf_y}{\omega^3}\sin\theta_{xy} \ .
    \label{eq:tbreak_02}
\end{equation}
As we said previously, it is necessary to have two nonzero in-plane components with nonzero $\theta_{xy}$. Any out-of-plane component will not contribute. Thus, even if there is some nonzero $\theta_{ij}$, only if $\theta_{xy}\neq 0$ there will be gap openings for the surface states. The Floquet Hamiltonian for the surface states will then be
\begin{equation}
    \mathcal{H}_{\mathrm{F,S}}=\left(\bm{\sigma}\times\bm{p}_{\bot}\right)\cdot\hat{\bm{z}}+\frac{f_xf_y}{\omega^3}\sin\theta_{xy}~\sigma_z \ .
    \label{eq:tbreak_03}
\end{equation}
Therefore, by applying a periodic driving, we induce a mass term into $\mathcal{H}_{\mathrm{F,S}}$, which leads to quasienergy gaps
\begin{equation}
    \Delta = 2\frac{f_xf_y}{\omega^3}\sin\theta_{xy} \ .
    \label{eq:tbreak_04}
\end{equation}
This observation can already be made by considering the surface effective Hamiltonian in the first place~\cite{Kitagawa11}. Our calculation also shows that out-of-plane components do not lead to gap openings. Additionally, we may check if this equation fits the numerical solution of the problem. As it can be seen in figure~\ref{fig:gap}, the agreement with the numerical calculations is significant. 
As it can be seen, by driving the system with different polarizations, the gap can be dynamically altered from zero to $2f_xf_y/\omega^3$. This quasienergy gap is non-trivial, in the sense that it occurs due to breaking of time-reversal symmetry. As a result, if the interface is finite in one direction, edge states or hinge modes appear and the system becomes a higher order topological insulator~\cite{Schindler18}.

In the next two subsections, we shall be interested on the widening of the Dirac cones by linear and circularly polarized fields. We noted earlier that such widening cannot be due to $\delta\mathcal{H}^{1}$ and we need to consider $\delta\mathcal{H}^{2}$. After some manipulations, we find
\begin{equation}
    \delta\mathcal{H}^{2}= \frac{1}{\omega^4}\left[-|\bm{f}|^2\mathcal{H}_0+\sum_{ij}f_if_jp_i\alpha_j\cos\theta_{ij}\right] \ .
    \label{eq:20}
\end{equation}
We will now particularize to the case of linear and circularly polarized fields. 

\subsection{Linear polarization}

\begin{figure}[htb]
\centerline{\includegraphics[width=0.95\columnwidth]{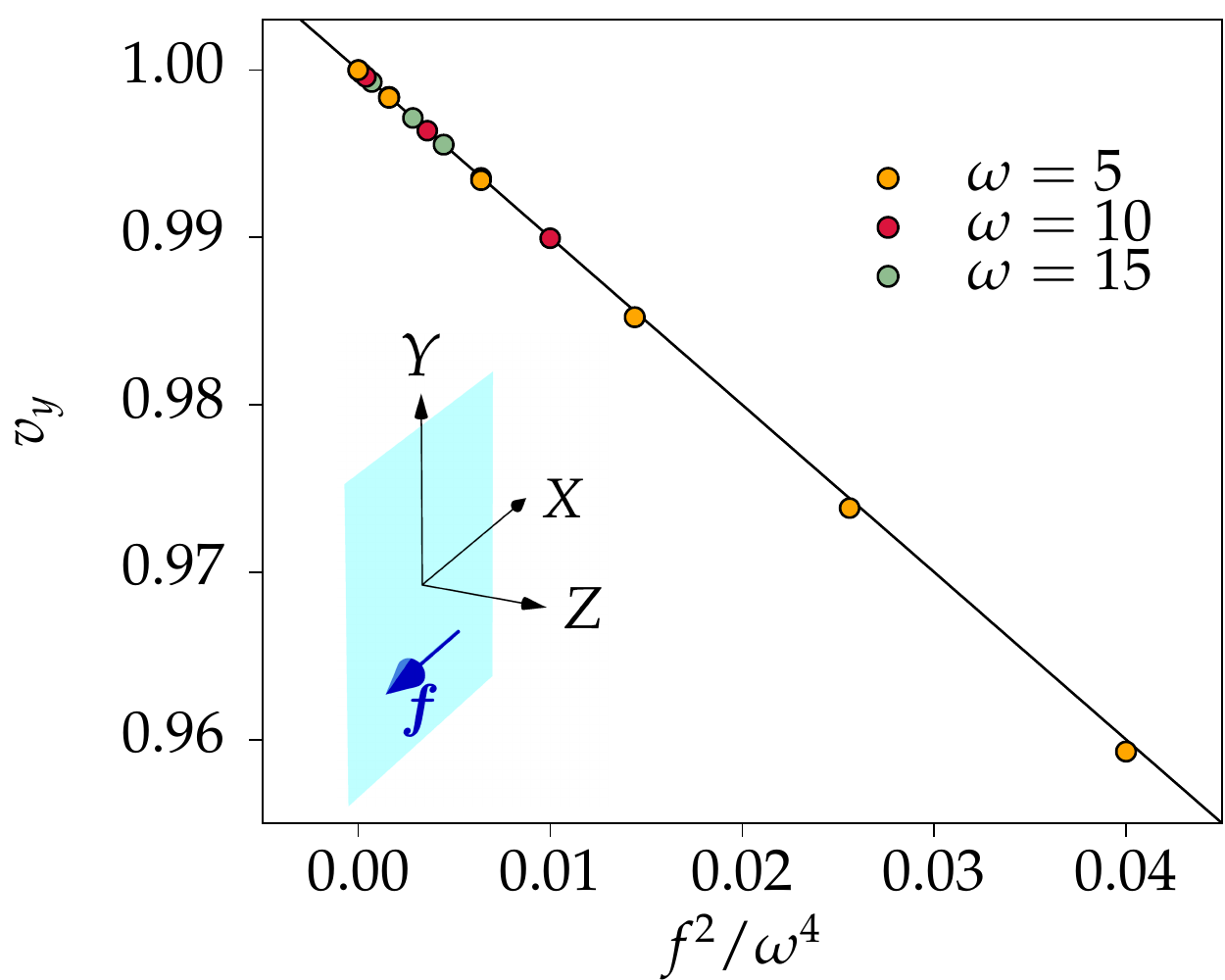}}
\caption{Reduction of the velocity along the $y$-direction as a function of $f^2/\omega^4$ for in-plane linearly polarized fields. The dots represent the numerical solution to the full problem and the black line corresponds to equation~(\ref{eq:27}).}
\label{fig:velocity_in_linear}
\end{figure}

We can see that if the polarization is linear, meaning that $\theta_{ij}=0$ for all $i,j$, then $\delta\mathcal{H}^1=0$ and the cosine term in $\delta\mathcal{H}^2$ equals one. The expression for $\delta\mathcal{H}^2$ simplifies even further if we consider two specific cases of linear polarization: in-plane and out-of-plane. In the former case, only one component of $\mathbf{f}$ contained in the interface is needed. Because the interface has continuous rotation symmetry, we can choose the component to point along any direction of our choice. Let us pick the $x$-direction for convenience. In that case, $f_x\equiv f$ and all other components of $\bm{f}$ are zero. In this situation, the Floquet Hamiltonian reads
\begin{equation}
    \mathcal{H}_{\mathrm{F}}\simeq\alpha_xp_x+\left(1-\frac{f^2}{\omega^4}\right)\left[\alpha_yp_y+\alpha_zp_z+\beta\sgn(z)\right] \ .
    \label{eq:22}
\end{equation}
which is nothing but the Hamiltonian for a topological boundary, equation~(\ref{eq:03}), if we rescale by a factor of $(1-f^2/\omega^4)^{-1}$ the $x$-component of the momentum and the quasienergy. Hence, we can conclude that the surface states survive under the external driving, their dispersions becoming anisotropic along the direction perpendicular to the driving, in this case the $y$-direction. Indeed, the velocities would simply be
\begin{equation}
    v_{x}=1 \ , \qquad v_{y} = 1-\frac{f^2}{\omega^4} \ .
    \label{eq:27}
\end{equation}
In figure~\ref{fig:velocity_in_linear} we compare the numerical result obtained from the procedure discussed in Ref.~\cite{Diaz-Fernandez19} with equation~(\ref{eq:27}) and the agreement is noteworthy. Moreover, we can numerically confirm that the surface states do not couple to the bulk states and, therefore, using surface effective Hamiltonians would have been appropriate in this case. One point that is not captured by the high-frequency expansion, at least not to second order, is the avoided crossings that take place at the Floquet-Brillouin zone edges, which can be observed in the numerical solution of the Hamiltonian~\cite{Diaz-Fernandez19} due to coupling between different Floquet side-bands. In any case, the anisotropic widening of the Dirac cones can be manipulated at will by properly adjusting the field and the frequency. 

\begin{figure}[htb]
\centerline{\includegraphics[width=0.95\columnwidth]{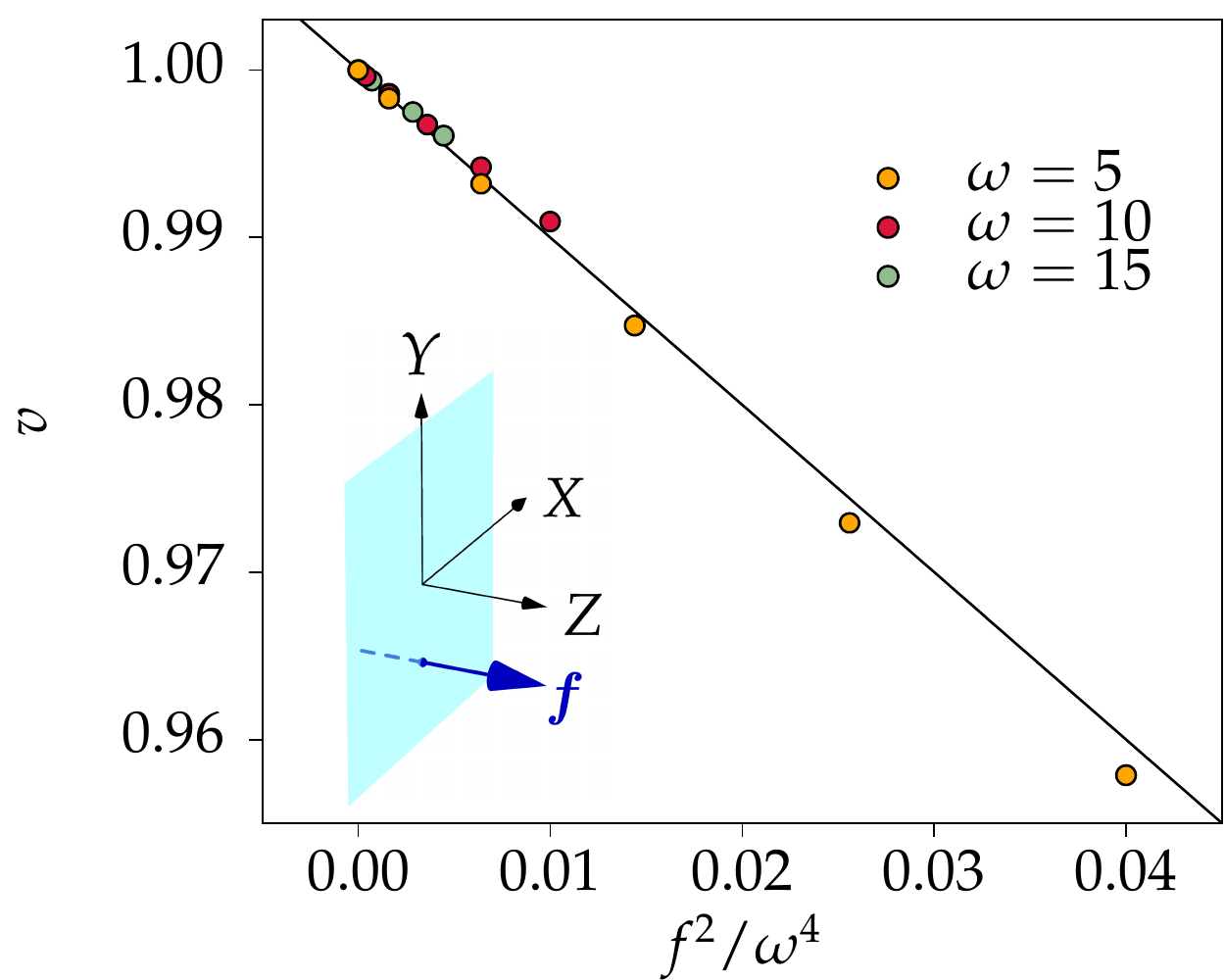}}
\caption{Reduction of the velocity as a function of $f^2/\omega^4$ for out-of-plane linearly polarized fields. The dots represent the numerical solution to the full problem and the black line corresponds to equation~(\ref{eq:34}).}
\label{fig:velocity_out_linear}
\end{figure}

Let us now turn to the case of having out-of-plane linearly polarized fields. In that case, $f_z\equiv f$ and all other components of $\bm{f}$ are zero. In that case, the Floquet Hamiltonian reads
\begin{equation}
    \mathcal{H}_{\mathrm{F}}\simeq \left(1-\frac{f^2}{\omega^4}\right)\left(\alpha_xp_x+\alpha_yp_y+\beta\sgn(z)\right)+\alpha_zp_z \ .
    \label{eq:28}
\end{equation}
As in the previous case, this is the Hamiltonian for a topological boundary with rescaled quasienergies by a factor of $(1-f^2/\omega^2)^{-1}$. However, in contrast to the previous case, the in-plane momenta are not rescaled but the $z$-coordinate is by a factor of $(1-f^2/\omega^2)$, which implies that the surface states become more delocalized as $f^2/\omega^4$ increases. In any case, for sufficiently small fields and high frequencies, the surface states remain localized but display an isotropic widening of the Dirac cone where the velocity is given by
\begin{equation}
    v = 1-\frac{f^2}{\omega^4} \ .
    \label{eq:34}
\end{equation}
A similar reduction has been observed in the static case theoretically~\cite{Diaz-Fernandez17a,Tchoumakov17} and experimentally~\cite{Inhofer17}. These observations are consistent with the numerical calculations. In particular, for the velocity we observe a noticeable agreement in figure~\ref{fig:velocity_out_linear}.
As we can see, the agreement starts to fail at large field amplitudes. This can be explained from the fact that the surface states become less localized as the field increases and their interaction with the bulk states becomes non-negligible. As a result, hybridizations with the bulk states can occur for surface states closer to the Floquet-Brillouin zone edges\cite{Diaz-Fernandez19}.

\subsection{Circular polarization}

\begin{figure}[htb]
\centerline{\includegraphics[width=0.95\columnwidth]{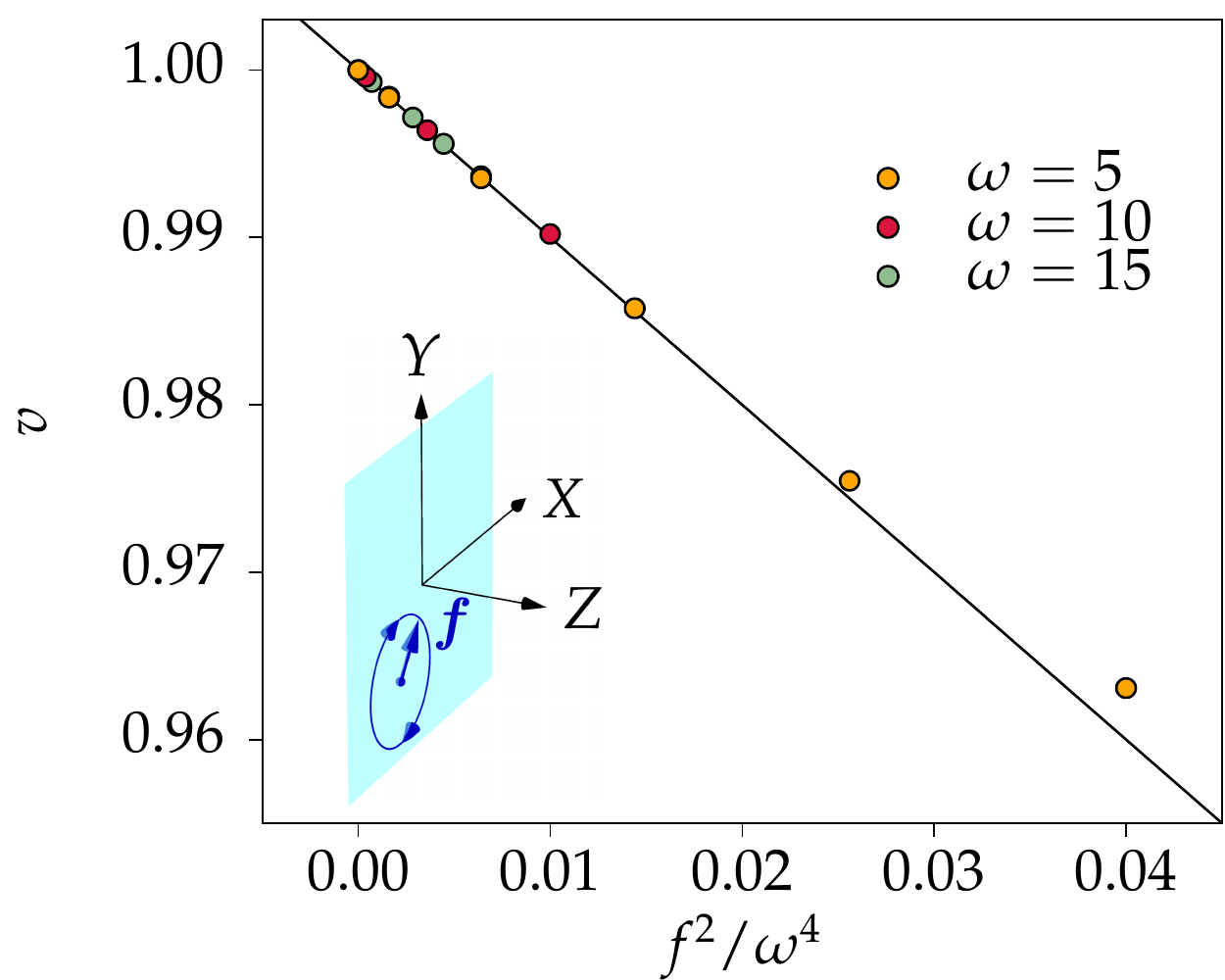}}
\caption{Reduction of the velocity as a function of $f^2/\omega^4$ for in-plane circularly polarized fields. The dots represent the numerical solution to the full problem and the black line corresponds to equation~(\ref{eq:40}).}
\label{fig:velocity_in_circular}
\end{figure}

For circular polarization, $\delta\mathcal{H}^1$ is no longer zero as we already discussed. We shall be interested in two cases: in-plane fields and out-of-plane fields. For the in-plane field, we have $f_x=f_y\equiv f$, $f_z=0$ and $\theta_{xy}=\pi/2$. Notice that the field amplitude, $|\bm{f}|$, will be a factor of $\sqrt{2}$ larger than in the in-plane case since $|\bm{f}|=\sqrt{2}f$ in this case. It is therefore important not to confuse $f$ with $|\bm{f}|$. In this situation, the Floquet Hamiltonian can be written as
\begin{equation}
\begin{aligned}
    \mathcal{H}_{\mathrm{F}} \simeq &  \left(1- \frac{2f^2}{\omega^4}\right)\left(\alpha_zp_z+\beta\sgn(z)\right) \\
    & + \left(1- \frac{f^2}{\omega^4}\right)(\bm{\alpha}\cdot\bm{p}_{\bot})+\frac{\Delta}{2}~\sigma_z \ ,
\end{aligned}
\label{eq:35}
\end{equation}
with $\Delta=2f^2/\omega^3$ as defined in equation~(\ref{eq:tbreak_04}) for $\theta_{xy}=\pi/2$. After rescaling, equation~(\ref{eq:35}) is that of a topological boundary with an additional mass term which, as discussed earlier, opens up a gap for the surface states at $\bm{p}_{\bot}=\bm{0}$. Additionally, there is an isotropic velocity reduction
\begin{equation}
 v = 1- \frac{f^2}{\omega^4} \ .
 \label{eq:40}
\end{equation}
The reduction is isotropic because of the symmetry of the driving, that is, because the driving is circular. However, as we have seen, a linear driving leads to anisotropic velocities. Therefore, for elliptical polarizations we would expect to have anisotropic velocities as well. Indeed, in the most general case of in-plane fields we would have for the quasienergy dispersion
\begin{equation}
\begin{aligned}
 \varepsilon^2=&  
 \left[\left(1-\frac{f_y^2}{\omega^4}\right)p_x+p_y\frac{f_xf_y}{\omega^4}\cos\theta_{xy}\right]^2 \\
 & \left[\left(1-\frac{f_x^2}{\omega^4}\right)p_y+p_x\frac{f_xf_y}{\omega^4}\cos\theta_{xy}\right]^2 \\
 & + \left(\frac{f_xf_y}{\omega^3}\sin\theta_{xy}\right)^2 \ .
 \end{aligned}
 \label{eq:disp_ellip}
\end{equation}
As it can be seen, the analysis is more complicated, but we can see that the dispersion can be modulated almost at will by considering different values for $f_x,f_y$ and $\theta_{xy}$. If $\theta_{xy}=\pi/2$ we have an anisotropic reduction
\begin{equation}
    v_x = 1-\frac{f_y^2}{\omega^4} \ , \qquad v_y=1-\frac{f_x^2}{\omega^4} \ .
    \label{eq:anisot_vel}
\end{equation}
The linearly polarized case requires one of the amplitudes to be zero as we have set $\theta_{xy}=\pi/2$ and we get the strongest anisotropy where the velocity changes only in the direction perpendicular to the field. The circularly polarized is isotropic as we discussed, but all cases in between are anisotropic as we were aiming for. We have already seen that the quasienergy gap perfectly matches the numerical calculations in figure~\ref{fig:gap}. We can also test if the reduction of the velocity fits the numerical calculations. Indeed it does, as shown in figure~\ref{fig:velocity_in_circular}.

\begin{figure}[htb]
\centerline{\includegraphics[width=0.95\columnwidth]{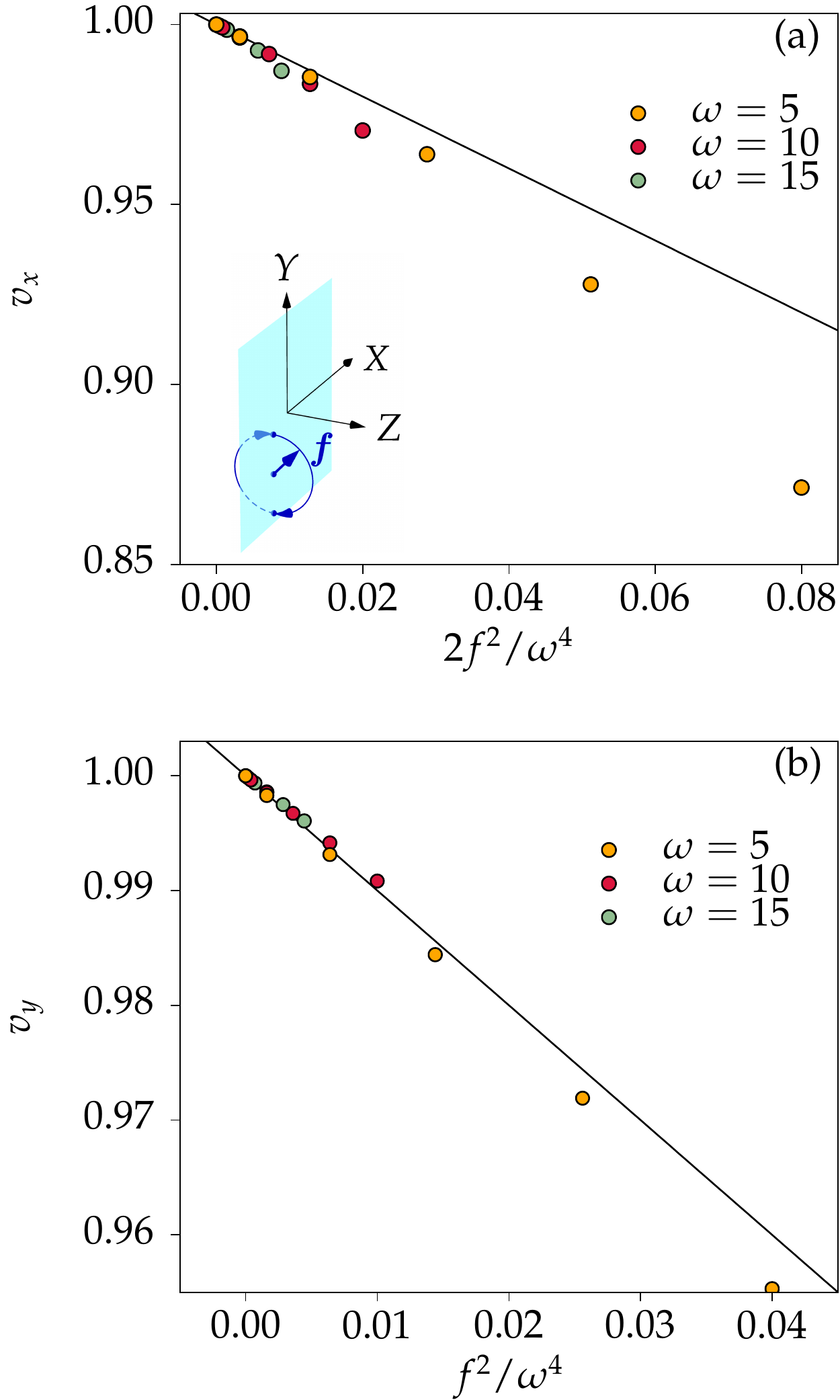}}
\caption{Reduction of the velocity along the (a) $x$-direction as a function of $2f^2/\omega^4$ and (b) $y$-direction as a function of $f^2/\omega^4$ for out-of-plane circularly polarized fields. The dots represent the numerical solution to the full problem and the black line corresponds to equation~(\ref{eq:47}).}
\label{fig:velocity_out_circular}
\end{figure}

Let us conclude this section with the last case, out-of-plane circularly polarized fields. We will choose $f_x=0$ and $f_y=f_z\equiv f$ and $\theta_{yz}=\pi/2$. In this situation, the Floquet Hamiltonian reads
\begin{equation}
\begin{aligned}
    \mathcal{H}_{\mathrm{F}} \simeq &  \left(1-\frac{2f^2}{\omega^4}\right)\left(\alpha_xp_x+\beta\sgn(z)\right) \\
    & +\left(1-\frac{f^2}{\omega^4}\right)\left(\alpha_yp_y+\alpha_zp_z\right) +\frac{\Delta}{2}\sigma_x 
\end{aligned}
\label{eq:41}
\end{equation}
with $\Delta=f^2/\omega^3$. This situation shares features with the linearly polarized in-plane and out-of-plane fields, since it can be thought of as the application of two independent linearly polarized fields in both directions. Indeed, the Hamiltonian above corresponds to that of a topological boundary after proper rescaling, where again the surface states become more delocalized and there is rescaling of $p_x$. Additionally, the mass term does not open a quasienergy gap as we already discussed. Regarding the anisotropic Dirac cone, we can already give an answer before turning to the actual results. We know that the in-plane component of the field leads to a reduction of the velocity in the direction perpendicular to the field [see equation~(\ref{eq:27})]. The out-of-plane component leads to isotropic reduction as shown by equation~(\ref{eq:34}). Therefore, we expect to observe a reduction of the velocity in both directions due to the out-of-plane component, with an enhanced reduction along the direction perpendicular to the in-plane component. This is precisely what we observe
\begin{equation}
    v_x = 1-\frac{2f^2}{\omega^4} \ , \quad v_y = 1-\frac{f^2}{\omega^4} \ .
    \label{eq:47}
\end{equation}
The question is how well does this fit the numerical results. As it can be seen in figure~\ref{fig:velocity_out_circular}, the agreement is not as good as in the previous cases considered in the paper. In particular, there is very little agreement in figure~\ref{fig:velocity_out_circular}(a) for large fields. Qualitatively, however, we do observe that the velocity is further reduced along the $x$-direction as compared to the $y$-direction. The strong deviations are due to the fact that the surface states hybridize with the bulk states the farther away we move from the Dirac point~\cite{Diaz-Fernandez19}. 

\section{Conclusions}

Topological insulators display Dirac cones at the surface which are protected by time-reversal symmetry. The transport properties associated to these cones are very distinct from ordinary semiconductors~\cite{Culcer12,Trescher15}. It is therefore interesting to be able to manipulate and reshape these cones, for instance by inducing anisotropies and opening non-trivial energy gaps. One such way to achieve this is by means of periodic modulation~\cite{Kitagawa11,Usaj14,Syzranov13}. In this article, we have considered a topological boundary as described by a three-dimensional Dirac Hamiltonian with a mass term which changes sign upon crossing the boundary. Using numerical calculations like those discussed in Ref.~\onlinecite{Diaz-Fernandez19} and high-frequency expansions up to second order, we have been able to predict the appearance of gap openings in the quasienergy spectrum when the external field has two nonzero in-plane components with nonzero phase difference. The results are in agreement to those obtained by means of surface effective Hamiltonians in the case of in-plane fields~\cite{Kitagawa11}. Additionally, the second order term in the expansion has allowed us to find analytical expressions for the widening of the cones, which occurs both isotropically and anisotropically, with very good agreement to the numerical calculations. By means of Floquet engineering, we have been able to show that such anisotropies in the Dirac cones can be achieved dynamically by simple external means, namely by changing the field orientations and the polarization. These anisotropies could potentially have an important impact in quantum transport as they do in the static case~\cite{Trescher15}. 

\acknowledgments

The author thanks F.~Dom\'{i}nguez-Adame for careful reading of the manuscript. This research has been supported by MINECO (Grant MAT2016-75955).

\appendix* 
\section{Floquet theory}

Let $\mathcal{H}(t)=\mathcal{H}(t+T)$ be a time-dependent periodic Hamiltonian with period $T$. Then Floquet theorem states that its eigenstates can be written as
\begin{equation}
 \Psi(t) = e^{-\ii \varepsilon t}\Phi(t) \ ,
    \label{eq:07}
\end{equation}
where $\Phi(t)=\Phi(t+T)$. In other words, $\Psi(t)$ is an eigenstate of the evolution operator over a single period $\mathcal{U}(T)$ with eigenvalue $\exp(-\ii\varepsilon T)$. As a result, $\varepsilon$ and $\varepsilon+n\omega$ describe the same state, with $n\in\mathbb{Z}$ and $\omega=2\pi/T$ the driving-frequency. Hence, in analogy to the quasimomentum of Bloch's theorem, $\varepsilon$ is called the quasienergy and it is restricted to the first Floquet-Brillouin zone, $\varepsilon\in[-\Omega/2,\Omega/2)$. $\Phi(t)$ satisfies
\begin{equation}
 \left[\ii\partial_t+\varepsilon\right]\Phi(t) = \mathcal{H}(t)\Phi(t) \ .
    \label{eq:08}
\end{equation}
The periodicity of $\Phi(t)$ allows us to Fourier expand it as
\begin{equation}
 \Phi(t) = \sum_{m}e^{-\ii m\omega t}\varphi_m \ .
    \label{eq:09}
\end{equation}
As a result
\begin{equation}
 (\varepsilon+n\omega)\varphi_n = \sum_{m}\mathcal{H}_{m}\varphi_{n-m}
    \label{eq:10}
\end{equation}
where $\mathcal{H}_m$ is the $m$-th Fourier component of $\mathcal{H}(t)$.

Because $\Psi(t)$ is an eigenstate of $\mathcal{U}(T)$ with eigenvalue $\exp(-\ii\varepsilon T)$, $\mathcal{U}(T)$ is commonly written as~\cite{Rudner20}
\begin{equation}
 \mathcal{U}(T) \equiv e^{-\ii \mathcal{H}_{\mathrm{F}}T} \ ,
    \label{eq:11}
\end{equation}
where $\mathcal{H}_{\mathrm{F}}$ represents an effective Floquet Hamiltonian. It should be noted, however, that $\mathcal{H}_{\mathrm{F}}$ is not the true Hamiltonian of the system since it is defined modulo $\omega$. However, if the frequency of the driving is sufficiently high, it can be shown that the effective Hamiltonian captures the physics of the true Hamiltonian. In that same limit, an expansion in powers of $\omega^{-1}$ can be put forward. Up to second order, we can write
~\cite{Goldman14,Bukov15,Eckardt15}
\begin{equation}
\begin{aligned}
 \mathcal{H}_{\mathrm{F}}\simeq & \mathcal{H}_0+\delta\mathcal{H}^1+\delta\mathcal{H}^{2} \ , \\
 \delta\mathcal{H}^{1}  = & \frac{1}{\omega}\sum_{m>0}\frac{1}{m}\left[\mathcal{H}_{m},\mathcal{H}_{-m}\right] \ , \\
 \delta\mathcal{H}^{2}  = & \frac{1}{\omega^2}\Bigg(\sum_{m\neq 0}\frac{1}{2m^2}\left[\mathcal{H}_{-m},\left[\mathcal{H}_0,\mathcal{H}_{m}\right]\right] \\
 & + \sum_{\substack{m\neq 0 \\ m'\neq 0,m}}\frac{1}{3mm'}\left[\mathcal{H}_{-m'},\left[\mathcal{H}_{m'-m},\mathcal{H}_m\right]\right]\Bigg) \ .
 \end{aligned}
    \label{eq:12}
\end{equation}
Notice that $\mathcal{H}_0$ is simply the time-average of $\mathcal{H}(t)$, which implies that it only contains the dc part of $\mathcal{H}(t)$.

\end{document}